\documentclass[
reprint,
amsmath,amssymb,nofootinbib,
aps,prl,floatfix,longbibliography
]{revtex4}

\usepackage{graphicx}
\usepackage{hyperref}
\usepackage[final]{changes}
\usepackage[capitalise]{cleveref}
\usepackage{tikz} 
\usetikzlibrary{positioning}
\usetikzlibrary{calc}

\begin{document}
\title{Scaling and intermittency in turbulent flows of elastoviscoplastic fluids}
\author{Mohamed S. Abdelgawad\footnote{The two authors contributed equally}}
\author{Ianto Cannon$^*$}
\author{Marco E. Rosti\footnote{Corresponding author: marco.rosti@oist.jp}}
\affiliation{Complex Fluids and Flows Unit, Okinawa Institute of Science and Technology Graduate University, 1919-1 Tancha, Onna-son, Okinawa 904-0495, Japan}
\date{\today}

\begin{abstract}
Non-Newtonian fluids have a viscosity that varies with applied stress. Elastoviscoplastic fluids, the elastic, viscous and plastic properties of which are interconnected in a non-trivial way, belong to this category. We have performed numerical simulations to investigate turbulence in elastoviscoplastic fluids at very high Reynolds-number values, as found in landslides and lava flows, focusing on the effect of plasticity. We find that the range of active scales in the energy spectrum reduces when increasing the fluid plasticity; when plastic effects dominate, a new scaling range emerges between the inertial range and the dissipative scales. An extended self-similarity analysis of the structure functions reveals that intermittency is present and grows with the fluid plasticity. The enhanced intermittency is caused by the non-Newtonian dissipation rate, which also exhibits an intermittent behaviour. These findings have relevance to catastrophic events in natural flows, such as landslides and lava flows, where the enhanced intermittency results in stronger extreme events, which are thus more destructive and difficult to predict.
\end{abstract}

\pacs{Valid PACS appear here}

\maketitle

\section{Introduction}
Many fluids in nature and industry exhibit a non-linear relationship between shear stress and shear rate, which is referred to as non-Newtonian behaviour. Several non-Newtonian features can exist, and they are often present simultaneously. Here, we focus on the so-called elastoviscoplastic (EVP) fluids, which are fluids with elastic, viscous, and plastic properties. EVP materials combine solid-like behaviour and fluid-like response depending on the value of the applied stress: they behave like a solid when the applied stress is below a critical value known as the \textit{yield stress}, and flow like a liquid otherwise~\cite{Balmforth2014}. The elastic nature of these materials is present in their solid as well as liquid states~\cite{Fraggedakis2016}. 
Such fluids are common in everyday life (e.g.~toothpaste, jam, cosmetics, mud), and turbulent flows of EVP fluids are found in many industrial processes, including sewage treatment, crude oil transportation, concrete pumping, and mud drilling~\cite{Hanks1963TheLT,10.2118/1682-PA,MALEKI201819}, and they are also found in nature as landslides and lava flows~\cite{Jerolmack2019,Jones2019}.

\begin{figure}[b!]
	\centering
			\includegraphics[width=0.49\textwidth]{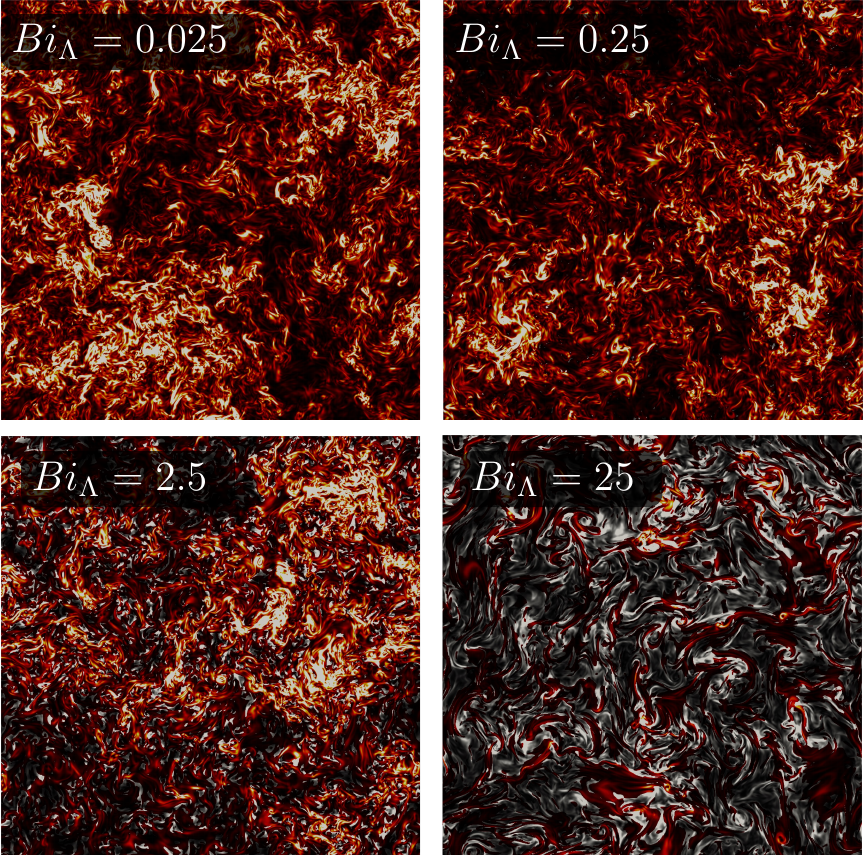}
	\caption{
Instantaneous colourmaps of the turbulent fluid dissipation $\epsilon_\text{f}$ in homogeneous isotropic turbulence of an EVP fluid at different Bingham numbers. Yielded regions are shown with the black-red-yellow colourscale, while unyielded regions with the black-gray-white one.
	}
	\label{fig:figure1}
\end{figure}

A great deal of work has been done in the past to properly characterize the viscoelastic behaviour of a fluid in both laminar and turbulent flows~\cite{Groisman2000,PoolePRL2007, Haward2016,V.Steinberg2021Rev,visco, abreu}, while the effect of plasticity has been studied mainly in low-Reynolds number laminar conditions~\cite{Pavlov1974HydrodynamicSO,ESCUDIER_JNNFM2005,Balmforth2014}. Little is known about the plastic behaviour of an EVP fluid in turbulence; \citet{Rosti2018} studied for the first time a turbulent channel flow of an EVP fluid, finding that the shape of the mean velocity profile controls the regions where the fluid is unyielded, forming plugs around the channel centreline that grow in size as the yield stress increases, similar to what is observed in a laminar condition. However, the presence of the plug region has an opposite effect on drag for laminar and turbulent flow configurations, resulting in drag reduction in the turbulent case and drag increase in the laminar one; the turbulent drag behaviour is due to the tendency of the turbulent flow to relaminarize, overall leading to a strongly non-linear relation between yield stress and drag coefficient. The simulation results were then employed by \citet{LeClainche2020} using high-order dynamic mode decomposition to study the near-wall dynamics, comparing them to those in Newtonian and viscoelastic fluids. Their work revealed that both elasticity and plasticity have similar effects on the near-wall coherent structures, where the flow is characterized by long streaks disturbed for short periods by localized perturbations. 
A recent experimental study by \citet{Mitishita2021FullyTF} on a turbulent duct flow of Carbopol solution de-facto verified the numerical results obtained by \citet{Rosti2018} on the effect of plasticity on the mean flow profile and Reynolds stresses. Additionally, they observed an increase in the energy content at large scales and a decrease at small scales, when compared to a Newtonian fluid. Mitishita et al. reported a $-7/2$ scaling in the energy spectra at high wavenumbers during Carbopol flows compared to $-5/3$ scaling in the case of water flows. The newly observed scaling was attributed either to the decrease in the inertial effect in the presence of Carbopol solutions that shrinks the inertial range of scales, since the Reynolds numbers are much lower than in water flows,  or to the elastic effects that become significant at large wavenumbers where the fluid experiences high frequencies. Moreover, the shear-thinning effects that Carbopol solutions exhibit affected the anisotropy and the overall flow behaviour. The elastic and shear thinning effects are rheological features of Carbopol solutions and can not be eliminated experimentally.

Homogeneous and isotropic turbulent flows have long been a focus of turbulence research for their simple theoretical analysis and the generality of their results. 
To this end, as has been extensively done in the past for viscoelastic flows, we study the tri-periodic homogeneous flow, where the celebrated K41 theory by Kolmogorov~\cite{kolmogorov41}, can be directly applied to a classical Newtonian fluid. In this work, we study for the very first time a homogeneous isotropic turbulent (HIT) flow of an elastoviscoplastic fluid at high Reynolds number, as shown in Fig. \ref{fig:figure1}. We aim to answer the following fundamental question: how does the Kolmogorov theory change when the fluid is elastoviscoplastic? We will mainly focus on its plastic behaviour and investigate how the yield stress affects the multiscale energy distribution and balance, and how the turbulent energy cascade is altered by the fluid's plasticity. Our results show profound modifications of the classical picture predicted by the K41 theory for Newtonian fluids, with the emergence of a new scaling range, the dominance of the non-Newtonian flux and dissipation at small and intermediate scales, and enhanced intermittency of the flow. 

\begin{figure}[t]
	\centering
	\includegraphics[width=0.49\textwidth]{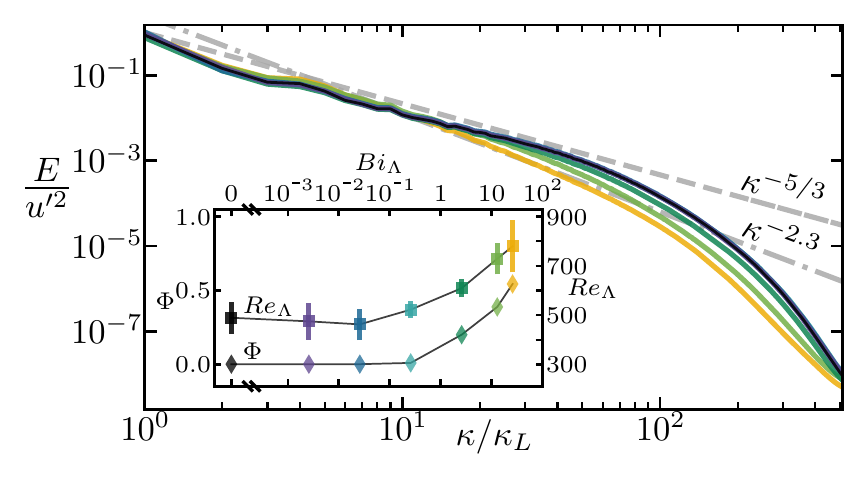}
	\caption{
Turbulent kinetic energy spectra of EVP flows with various Bingham numbers, plotted in colours from dark to light; $Bi_\Lambda=0, 0.0025,~0.025,~0.25,~2.5,~12.5$ and 25 are plotted in black, purple, dark blue, light blue, dark green, light green, and orange respectively. The expected Kolmogorov scaling for a Newtonian fluid is shown by a grey dashed line, while the dash-dotted line represents an apparent new non-Newtonian scaling $E\sim \kappa^{-2.3}$ which emerges at large $Bi_\Lambda$. The inset of the figure reports how the mean values of the micro-scale Reynolds number $Re_\Lambda$ (plotted using squares on the right axis) and the volume fraction of the unyielded regions $\Phi$ (plotted using diamonds on the left axis) vary as a function of $Bi_\Lambda$. Error bars report the standard deviation of $Re_\Lambda$ in time, measured using $10^3$ samples. Plastic effects start to appear for $Bi_\Lambda\gtrsim 1$, suggesting that $\Lambda$ is the relevant length scale of the problem.
}
	\label{fig:figure2}
\end{figure}

\begin{figure}[t]
	\centering
	\includegraphics[width=0.49\textwidth]{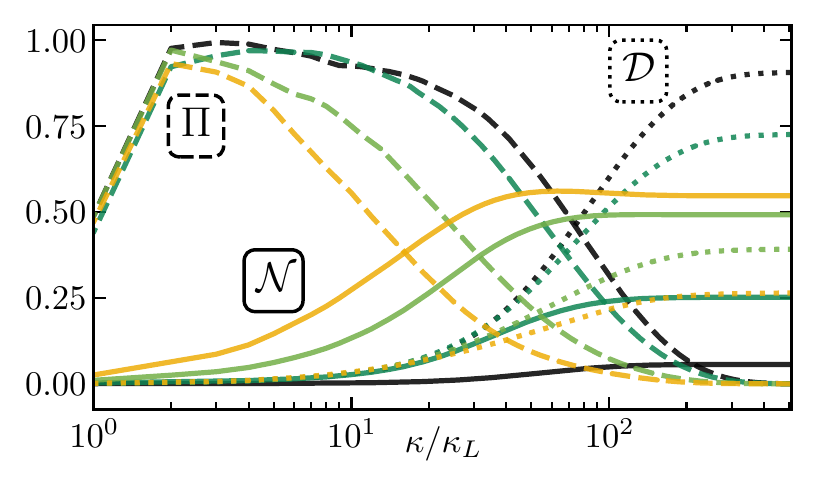}
	\caption{
Scale-by-scale energy balance for $Bi_\Lambda=0$ in black, $Bi_\Lambda=2.5$ in dark green, $Bi_\Lambda=12.5$ in light green, and $Bi_\Lambda=25$ in orange. We plot the energy flux of the non-linear convective term $\Pi$ using dashed lines, the solvent dissipation $\mathcal{D}$ using dotted lines, and the non-Newtonian contribution $\mathcal{N}$ using solid lines. Each term is normalized by the total dissipation rate $\langle \epsilon_\text{t} \rangle $. $\mathcal{N}$ grows at intermediate and small scales when $Bi_\Lambda$ is increased, eventually becoming the dominant contribution.
}
	\label{fig:figure3}
\end{figure}

\section{Results}
To investigate the problem at hand, we perform massive three-dimensional direct numerical simulations (DNS) of HIT where we solve the flow equations fully coupled with the constitutive equation of the EVP fluid, within a tri-periodic domain of size $L$, using $1024$ grid points per side, as described in more detail in the Methods section.
The flow is controlled by four main parameters: the Reynolds number $Re_\Lambda$, the Weissenberg number $Wi_\Lambda$, the viscosity ratio $\alpha$, and the Bingham number $Bi_\Lambda$, all based on the root mean square velocity fluctuations $u'$ and Taylor's micro-scale $\Lambda$. We use the definitions {$Re_\Lambda \equiv \rho u' \Lambda / \mu _\protect \text {t}$, $Wi_\Lambda \equiv\lambda \mu _\protect\text{t}/\rho\Lambda_0^2$, $\alpha = \mu _\protect \text {n}/\mu _\protect \text {t}$, and $Bi_\Lambda  \equiv \tau _y \Lambda_0 /\mu _\protect \text {t} u_0'$, where $\rho$ is the fluid density, $\mu _\protect \text {t}\equiv\mu_\protect \text {f} + \mu _\protect \text {n}$} is the total dynamic viscosity with $\mu _\protect \text {f}$ being the fluid viscosity and $\mu _\protect \text {n}$ the non-Newtonian one, $\lambda $ is the relaxation time, and $\tau _y$ is the yield stress, and subscript 0 denotes quantities from the $Bi_\Lambda=0$ case.
The Reynolds number describes the ratio of inertial to viscous forces, and we limit our analysis to high-Reynolds number flows, achieving a Taylor micro-scale Reynolds number $Re_{\Lambda} \approx 435$ for the Newtonian flow, at which statistics of the flow have been found to be universal and exhibiting a proper scale separation, with an extensive inertial range of scales extended to almost two decades of wavenumbers. The Reynolds number explored here is the highest ever reached in DNS of HIT of non-Newtonian fluids. The Weissenberg number describes the ratio of elastic to viscous forces, and here we limit the analysis to $Wi_\Lambda \ll 1$, (i.e., $Wi_\Lambda \approx  {10^{-3}}$), to ensure that elastic effects are sub-dominant and all the observed changes are due to plasticity. We also fix the value of $\alpha= 0.1$ to represent a dilute concentration of polymers, in accordance with prior works on the subject~\cite{Perlekar2006, Rosti2018}. Thus, the key control parameter we vary is $Bi_\Lambda$, which describes the ratio of the yield stress to the viscous stress, and thus correlates with the prevalence of unyielded  regions. 

Fig.~\ref{fig:figure2} depicts the turbulent kinetic energy spectra of the cases analysed. The $Bi_\Lambda=0$ case is similar to the Newtonian case shown in Fig.~\ref{fig:Fig1S} of the Supplementary Information, confirming that the effect of elasticity is subdominant and can be ignored. A clear $E\sim\kappa^{-5/3}$ range is visible for more than one decade, showing $Re_\Lambda$ is high enough to achieve scale separation, with the spectra exhibiting an inertial range of scales followed by a dissipative range. As $Bi_\Lambda$ increases, the inertial range is limited to the large scales (small wavenumbers $\kappa$), with the energy increasing at the large scales while decreasing at the small scales. A clear deviation from the Kolmogorov scaling becomes noticeable for $Bi_\Lambda > 1$, resulting in the emergence of a new apparent scaling of $E \sim \kappa^{-2.3}$ that is shown more clearly by plotting compensated energy spectra (as shown in Fig.~\ref{fig:Fig3S} of the Supplementary Information). The difference in scaling between the experimental work ($-7/2$)~\cite{Mitishita2021FullyTF} and the current study ($-2.3$) is mainly due to the higher values of Reynolds and Bingham numbers considered here.
The abrupt change in the spectra with $Bi_\Lambda$ is consistent with the bulk flow properties ($Re_\Lambda$ and the volume fraction of the unyielded regions $\Phi$), shown in the inset of the figure: for the cases where $Bi_\Lambda < 1$, $Re_{\Lambda}$ remains relatively unaltered, with $\Phi$ always close to zero, whereas when $Bi_\Lambda$ further increases, the micro-scale Reynolds number $Re_\Lambda$ and the volume $\Phi$ of the unyielded regions rapidly increase with a similar trend.

To fully characterize the change in the energy spectra, we study the turbulent kinetic energy balance, which in wavenumber space can be expressed as
\begin{equation}
	\mathcal{F}_\text{inj}(\kappa) + \Pi (\kappa) + \mathcal{D}(\kappa)+\mathcal{N}(\kappa) = \langle \epsilon_\text{f} \rangle +\langle \epsilon_\text{n} \rangle = \langle \epsilon_\text{t} \rangle,
	\label{eq:energy-transfer}
\end{equation}
where $\mathcal{F}_{\text{inj}}$ is the turbulence production introduced by the external forcing (injected at the largest scale $\kappa_{L}\equiv2\pi/L$); $\Pi$, $\mathcal{D}$, and $\mathcal{N}$ are the non-linear energy flux, the fluid dissipation, and the non-Newtonian contribution, respectively. In addition to the classical bulk fluid dissipation rate $\epsilon_\text{f}$, here we have a non-Newtonian dissipation $\epsilon_\text{n}$ which 
is the rate of removal of turbulent kinetic energy from the flow due to the non-Newtonian extra stress tensor (see the Supplementary Information for a derivation of this equation). Fig. \ref{fig:figure3} shows the turbulent kinetic energy balance for a few representative values of $Bi_\Lambda$. When comparing with Fig.~\ref{fig:Fig1S}b of the Supplementary Information, the $Bi_\Lambda =0$ case closely follows the classical Newtonian turbulent flow, wherein energy is carried by $\Pi$ from the large to small scales before being dissipated by the fluid viscosity $\mathcal{D}$. The contribution of the non-linear convective term $\Pi$, which appears as an almost horizontal plateau at relatively large scales, progressively decreases with $Bi_\Lambda$ and shrinks towards larger scales, consistently with the reduction of the extension of the inertial range observed in Fig. \ref{fig:figure2}. The reduced energy flux with $Bi_\Lambda$ is also accompanied by a decrease of the fluid dissipation $\mathcal{D}$, which are instead compensated by the increase of non-Newtonian contribution $\mathcal{N}$. At the small scales (large $\kappa$), the relative importance of the non-Newtonian contribution increases with $Bi_\Lambda$, becoming comparable to the fluid dissipation for $Bi_\Lambda \approx 2.5$ and eventually becoming the dominant term for $Bi_\Lambda \gtrsim 12.5$, corresponding to the emergence of the new scaling in the energy spectrum shown in Fig. \ref{fig:figure2}; indeed, the non-Newtonian contribution can be interpreted as a combination of a pure energy flux (giving rise to the new scaling region) and a pure dissipative term, as recently suggested by \citet{Rosti2021}. Regarding the direction of energy flux, Fig.~\ref{fig:Fig4S_S3ByFluidDissVsR} in the Supplementary Information shows that we have a direct cascade of energy from large to small scales for all $Bi_\Lambda$~\cite{xia_upscale_2011-1,cerbus_third-order_2017}.
\begin{figure}[t]
	\centering
	\includegraphics[width=0.99\textwidth]{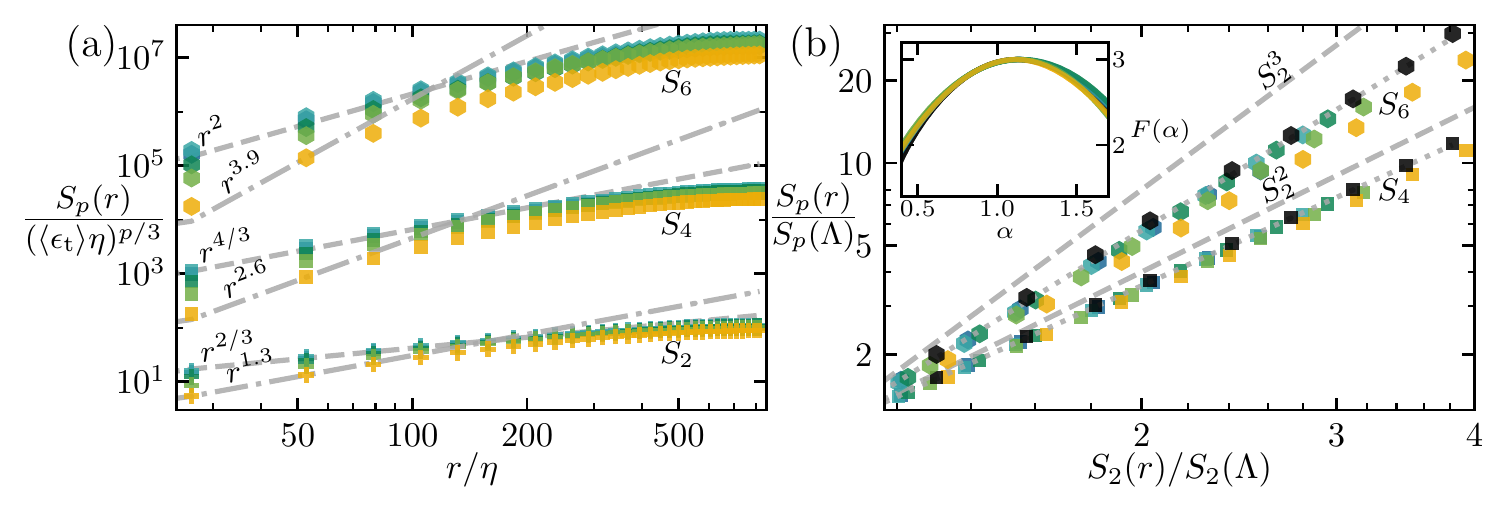}
	\caption{
(a): Dependence of the longitudinal velocity structure functions $S_2$ (pluses), $S_4$ (squares), and $S_6$ (hexagons) on the separation distance $r$. Symbol colour denotes $Bi_\Lambda$ and is the same as in Fig.~\ref{fig:figure2}. Dashed lines show scalings predicted by K41, and dash-dotted lines show scalings predicted using the new non-Newtonian scaling $E\sim \kappa^{-2.3}$. (b): The two scalings collapse onto a single line when we plot the structure functions $S_4$ and $S_6$ in extended self-similarity form, i.e., against $S_2$. To easily see changes in gradient, we have normalized the structure functions by their values at $r \approx \Lambda$. The dotted line shows a best fit through the data for $Bi_\Lambda= 0$, which deviates from the K41 prediction (dashed line) due to intermittency. Increasing $Bi_\Lambda$ further increases this deviation. The inset in (b) reports the multifractal spectrum of the energy dissipation rate carried out by the fluid $\epsilon_\text{f}$. 
}
	\label{fig:figure4}
\end{figure}

We extend the analysis done in the spectral domain, by computing the longitudinal structure functions defined as $S_{\text{p}}(r)= \langle (\Delta u(r))^{p} \rangle$, where $p$ is the order of the structure function and $\Delta u(r)=u(x+r) - {u}(x)$ is the difference in the fluid velocity across a length scale $r$, projected in the direction of $r$. According to K41, $S_{\text{p}}(r)~\sim \left( \langle \epsilon_\text{t} \rangle r \right)^{p/3}$; however, when the structure functions are displayed as a function of $r$, as shown in Fig. \ref{fig:figure4}a, they deviate from the K41 prediction as $p$ increases. This phenomenon is thought to be due to the intermittency of the flow, i.e., extreme events which are localised in space and time, and thereby break Kolmogorov's hypothesis of self similarity in the inertial range~\cite{Frisch}. Intermittency results in the scaling exponent of $r$ being a non-linear concave function of $p$ (instead of $p/3$)~\cite{kolmogorov_1962}. For the EVP fluid, two scaling regions appear at large $Bi_\Lambda$, with scaling consistent with those from the energy spectra, and with intermittency present in both scaling regions. The role of intermittency in the scaling exponents can be better appreciated when the structure functions are displayed in their extended self-similarity form, obtained by plotting one structure function against another~\cite{PhysRevE.48.R29}. In Fig. \ref{fig:figure4}b, $S_4$ and $S_6$ are plotted against $S_2$ for all Bingham numbers considered. We note a clear power-law scaling, which deviates from Kolmogorov's prediction, even for the $Bi_\Lambda=0$ case shown in black.
The departure from Kolmogorov's prediction progressively grows as the plasticity of the fluid increases, suggesting that the flow becomes more intermittent due to its plasticity. This becomes more obvious when we plot $S_n$ compensated by the intermittency correction at $Bi_\Lambda=0$ against $S_2$ (see Fig.~\ref{fig:Fig5S} in the Supplementary Information). Also, intermittency appears to act equally in the two scaling regions present at large $Bi_\Lambda$.

Intermittency originates from the multifractal nature of the turbulent dissipation rate~\cite{Frisch}. For Newtonian fluids, this can be quantified by the multifractal spectrum of the energy dissipation rate $\epsilon_\text{f}$~\cite{mandelbrot_1974, Frisch}, which we report in the inset of Fig. \ref{fig:figure4}b. The graph demonstrates that $F(\alpha)$ is nearly identical for all $Bi_\Lambda$ cases except for minor variations at small and large values of $\alpha$. This implies that the fluid dissipation rate is not the cause of the enhanced intermittency observed in the extended self-similarity analysis.

\begin{figure}[t]
	\centering
	\includegraphics[width=0.99\textwidth]{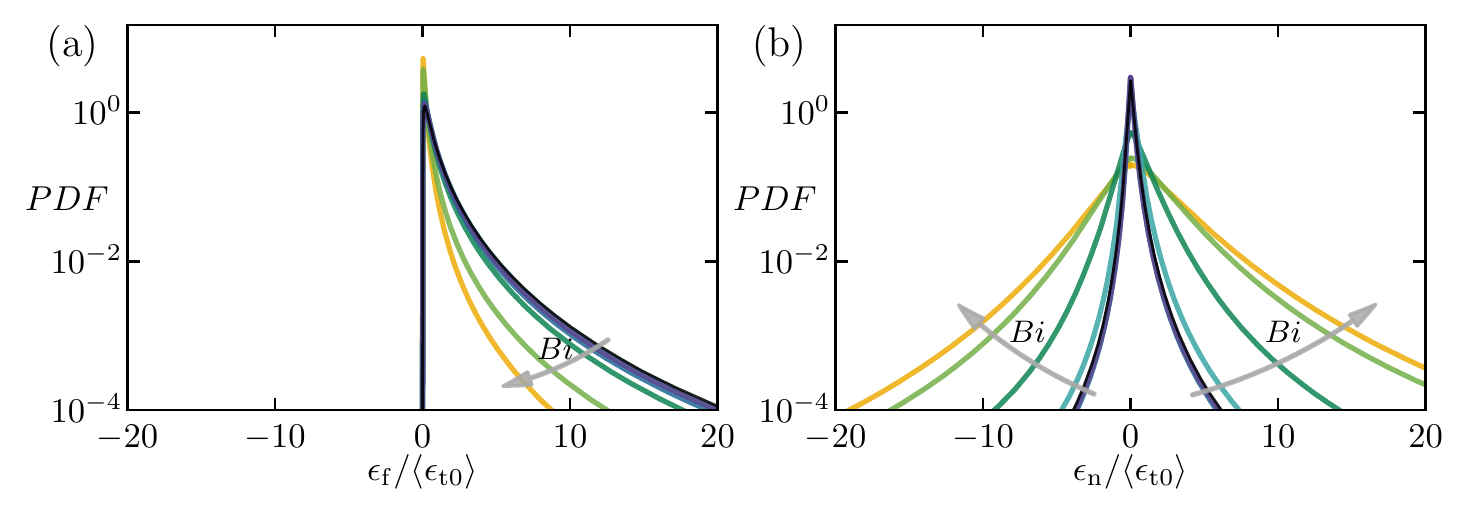}
	\caption{
Probability distribution function (PDF) of (a) the fluid dissipation rate $\epsilon _\text{f}$ and of (b) the non-Newtonian dissipation rate $\epsilon _\text{n}$ averaged over time, and normalised by $\langle \epsilon_0 \rangle$ the total dissipation of the $Bi_\Lambda=0$ flow. As $Bi_\Lambda$ increases, the PDF of the fluid dissipation rate $\epsilon _\text{f}$ slightly narrows, while the PDF of the non-Newtonian contribution $\epsilon _\text{n}$ significantly widens.
}
	\label{fig:figure5}
\end{figure}

In the present flow, the turbulent kinetic energy is dissipated by two different terms $\epsilon_\text{f}$ and $\epsilon_\text{n}$ seen in Fig. \ref{eq:energy-transfer}; hence, we investigate their respective behaviour by looking at their probability distribution functions in Fig. \ref{fig:figure5}. We name the non-Newtonian contribution $\epsilon _\text{n}$ a ``dissipation'' because on average it removes energy from the flow, giving rise to the positive-skewed distributions in Fig. \ref{fig:figure5}b; however, unlike the fluid dissipation, it can take positive or negative values at particular locations in space and time. Fig. \ref{fig:figure5}a shows that the distribution of $\epsilon _\text{f}$ narrows as $Bi_\Lambda$ increases~\cite{donzis_scalar_2005}; on the other hand, from Fig. \ref{fig:figure5}b, we see that that the distribution of $\epsilon_\text{n}$ significantly broadens as $Bi_\Lambda$ increases. Since the non-Newtonian dissipation becomes dominant for the largest $Bi_\Lambda$ (as shown in Fig. \ref{fig:figure3}), we can thus infer that the extreme values of $\epsilon_\text{n}$ are indeed the source of the enhanced intermittency observed from the structure function analysis in Fig. \ref{fig:figure4}.

\section{Discussion}
By means of unprecedented high-Reynolds-number DNS of an elastoviscoplastic fluid, we have shown that plastic effects significantly alter the classical turbulence predicted by the Kolmogorov theory for Newtonian fluids.

We have proved that the non-Newtonian contribution to the energy balance becomes dominant at intermediate and small scales for large Bingham numbers, inducing the emergence of a new intermediate scaling range in the energy spectra between the Kolmogorov inertial and the dissipative ranges, where energy spectrum decays with a $-2.3$ exponent. Interestingly, this exponent has been recently found for turbulence of viscoelastic fluids at large Reynolds and Weissenberg number~\cite{Rosti2021,Zhang2021}, suggesting a possible similarity among plastic and elastic effects on the turbulent cascade. This similarity in the scaling behaviour of the two cases could be attributed to a similar interaction mechanism in the Navier-Stokes equation between the convective and extra stress terms. It is also worth noting that in the context of viscoelastic flows at high Weissenberg number, an exponent less than or equal to -$3$ has been widely reported in the past~\cite{Groisman2000}; however, this is only found at relatively lower Reynolds number than investigated here or explored in recent experimental and numerical work~\cite{Rosti2021,Zhang2021}. The present work appears to be the first to report the $-2.3$ scaling in turbulent flows of highly plastic EVP fluids, and further studies on the size and distribution of the unyielded regions could shed more light on the origin of the newly found scaling.

We have also shown that the flow in the presence of plastic effects is more intermittent than in a Newtonian fluid, due to the combination of the classical intermittency originating from the multifractal nature of the turbulent dissipation rate, which remains substantially unaltered, and a new plastic contribution which instead grows with the Bingham number. A direct consequence of this result is that intermittency corrections for an elastoviscoplastic fluid are non-universal and dependent of the flow configuration, differently from viscoelastic flows. These results are relevant for several catastrophic natural flows with high plasticity, e.g., lava flows and landslides ~\cite{schaeffer_steady_2008}. Our findings explain why such flows are usually found to be intermittent and frequently aggressive, resulting in more damage. The non-universality of the flow intermittency in elastoviscoplastic fluids reflects also in an increased difficulty in their modelling.

\section{Acknowledgments}
	The research was supported by the Okinawa Institute of Science and Technology Graduate University (OIST) with subsidy funding from the Cabinet Office, Government of Japan. The authors acknowledge the computer time provided by the Scientific Computing section of Research Support Division at OIST and the computational resources of the supercomputer Fugaku provided by RIKEN through the HPCI System Research Project (Project IDs: hp210229 and hp210269).

\section{Author contributions}
M.E.R. conceived the original idea, planned the research, and developed the code. All authors performed the numerical simulations, analysed data, outlined the manuscript content and wrote the manuscript.

\section{Competing interests}
The authors declare that they have no competing interests.

\section{Methods}

\subsection{Governing equations}

The flow under investigation is governed by a system of a scalar, a vector and a tensorial equation, these are the incompressibility constraint, the conservation of momentum, and the constitutive equation for the non-Newtonian extra stress tensor, respectively. The incompressibility constraint and the momentum conservation equations can be written as
\begin{equation}\label{eq:mass}
	\nabla \cdot \boldsymbol{u}=0,
\end{equation}

\begin{equation}\label{eq:mom}
	\rho\left(\frac{\partial \boldsymbol{u}}{\partial t}+\boldsymbol{u} \cdot\nabla \boldsymbol{u}\right) = \nabla p + \mu_\text{f} \nabla^2  \boldsymbol{u}+ \boldsymbol{f}_\text{inj} +\boldsymbol{f}_{\text{evp}},
\end{equation}
where $\boldsymbol{u}$ is the fluid velocity, $p$ is the pressure, $\rho$ is the density, and $\mu_\text{f}$ is the fluid dynamic viscosity. The term $\boldsymbol{f}_\text{inj}$ represents the external force used to sustain turbulence; here we consider the Arnold-Beltrami-Childress (ABC) flow with forcing 
\begin{equation}\label{eq:f_turb}
	\boldsymbol{f}_\text{inj} = \boldsymbol{i} \mu_\text{f} (A \sin z/L + C \cos y/L) 
	+	\boldsymbol{j} \mu_\text{f} (B \sin y/L + A \cos z/L)
	+	\boldsymbol{k} \mu_\text{f} (C \sin y/L + B \cos x/L),
\end{equation}
where $\boldsymbol{i}$, $\boldsymbol{j}$, $\boldsymbol{k}$ are the Cartesian unit vectors, $A$, $B$, and $C$ are real parameters, and the flow has periodicity $L$ in $x$, $y$, and $z$.
In our simulations, we choose $A=B=C$ and use an appropriate value of $\mu_\text{f}$ to give a micro-scale Reynolds number $\mathrm{Re}_\Lambda \approx 435$ for the Newtonian flow.
The last term in equation~\ref{eq:mom} is defined as $\boldsymbol{f}_{\text{evp}}\equiv \nabla \cdot \boldsymbol{\tau}$, where $\boldsymbol{\tau}$ is the non-Newtonian extra stress tensor of the EVP fluid.
We adopt the constitutive model proposed by Saramito~\cite{Saramito2007} to express the evolution of the extra stress tensor which can be written as 
\begin{equation}\label{eq:evp}
	\lambda {\overset{\nabla}{\boldsymbol{\tau}}}+\max\left(0,\frac{\tau_{d}-\tau_{y}}{\tau_{d}}\right) \boldsymbol{\tau}= \mu_\text{n} \left( \nabla\boldsymbol{u}+\left(\nabla\boldsymbol{u}\right)^{T} \right)
\end{equation}
where $(\overset{\nabla}{.})$ denotes the upper convected derivative, i.e.,  {$\overset{\nabla}{\boldsymbol{\tau}}=\frac{\partial\boldsymbol{\tau}}{\partial t}+\boldsymbol{u}\cdot\nabla\boldsymbol{\tau}-\boldsymbol{\tau}\cdot\nabla\boldsymbol{u}-(\nabla\boldsymbol{u})^{T}\cdot\boldsymbol{\tau}$.} 
$\mu_\text{n}$ is the non-Newtonian dynamic viscosity, $\tau_{d}$ is the magnitude of the deviatoric part of the stress tensor $\boldsymbol{\tau}_{d}\equiv \boldsymbol{\tau}- \mathrm{tr}(\boldsymbol{\tau})\boldsymbol{I}/3$, and $\boldsymbol{I}$ is the identity tensor, i.e., $\tau_{d}= \sqrt{\frac{1}{2}(\boldsymbol{\tau}_d:\boldsymbol{\tau}_d)}$. Before yielding, i.e., $\tau_d \le \tau_{y} $, the model predicts only recoverable Kelvin-Voigt viscoelastic deformation, while after yielding, i.e., $\tau_d > \tau_{y}$, it predicts Oldroyd-B viscoelastic behaviour. This transition occurs in a continuous manner. There are other EVP models that take into account shear-thinning~\cite{SARAMITO2009} or thixotropic behaviour~\cite{DIMITRIOU2019}; however, we chose the one described above for its simplicity and the least number of involved parameters. Also, this model proved able to capture the main flow characteristics in a turbulent channel flow \cite{Rosti2018, Mitishita2021FullyTF}.

\subsection{Numerical method}
\label{sec:numerical}
We use the in-house flow solver \textit{Fujin} (\url{https://groups.oist.jp/cffu/code}) to solve the governing equations numerically on a staggered uniform Cartesian grid; velocities are located on the cell faces, while pressure, stresses, and the other material properties are located at the cell centres. The second-order central finite difference scheme is used for spatial discretisation except for the advection term that comes from the upper convective derivative in \cref{eq:evp} where the fifth-order WENO (weighted essentially non-oscillatory) scheme is adopted~\cite{shu2009}. The second-order Adams-Bashforth scheme coupled with a fractional step method~\cite{kim_application_1985} is used for the time advancement of all terms except for the non-Newtonian extra stress tensor, which is advanced with the Crank-Nicolson scheme. To enforce a divergence-free velocity field, a fast Poisson solver based on the Fast Fourier Transform (FFT) is used for the pressure. The domain decomposition library \textit{2decomp} (\texttt{http://www.2decomp.org}) and the MPI protocol are used to parallelize the solver. The evolution equation of the extra EVP stress is formulated and solved using the log-conformation method~\cite{izbassarov_etal_2018a} to ensure the positive-definiteness of the conformation tensor. The fluid domain is a periodic cubic box of length $L$ discretized using $1024$ grid points per side, resulting in a large grid resolution sufficient to represent the fluid properties at all the scales of interest with $\eta/\Delta x = \mathcal{O}(1)$, where $\eta$ is the Kolmogorov length-scale, and $\Delta x$ is the grid spacing.

\section{Data availability}
All data needed to evaluate the conclusions are present in the paper and/or the Supplementary Information. The data that support the findings of this study are openly available in OIST at \url{https://groups.oist.jp/cffu/abdelgawad2023natphys}.

\section{Code availability}
The code used for the present research is a standard direct numerical simulation solver for the Navier--Stokes equations. Full details of the code used for the numerical simulations are provided in the Methods section and references therein.

\section{Methods references}
\begin{enumerate}
	\setcounter{enumi}{30} 
	\itemsep0em 
    \expandafter\ifx\csname natexlab\endcsname\relax\def\natexlab#1{#1}\fi
	\expandafter\ifx\csname bibnamefont\endcsname\relax
	\def\bibnamefont#1{#1}\fi
	\expandafter\ifx\csname bibfnamefont\endcsname\relax
	\def\bibfnamefont#1{#1}\fi
	\expandafter\ifx\csname citenamefont\endcsname\relax
	\def\citenamefont#1{#1}\fi
	\expandafter\ifx\csname url\endcsname\relax
	\def\url#1{\texttt{#1}}\fi
	\expandafter\ifx\csname urlprefix\endcsname\relax\def\urlprefix{URL }\fi
	\providecommand{\bibinfo}[2]{#2}
	\providecommand{\eprint}[2][]{\url{#2}}
\setcounter{enumiv}{9}	
	\bibitem[{\citenamefont{Saramito}(2007)}]{Saramito2007}
	\bibinfo{author}{\bibfnamefont{P.}~\bibnamefont{Saramito}},
	  \bibinfo{journal}{Journal of Non-Newtonian Fluid Mechanics}
	  \textbf{\bibinfo{volume}{145}}, \bibinfo{pages}{1} (\bibinfo{year}{2007}),
	  ISSN \bibinfo{issn}{03770257},
	  \urlprefix\url{https://linkinghub.elsevier.com/retrieve/pii/S0377025707000869}.
	
	\bibitem[{\citenamefont{Saramito}(2009)}]{SARAMITO2009}
	\bibinfo{author}{\bibfnamefont{P.}~\bibnamefont{Saramito}},
	  \bibinfo{journal}{Journal of Non-Newtonian Fluid Mechanics}
	  \textbf{\bibinfo{volume}{158}}, \bibinfo{pages}{154} (\bibinfo{year}{2009}),
	  ISSN \bibinfo{issn}{0377-0257},
	  \urlprefix\url{https://www.sciencedirect.com/science/article/pii/S0377025708002267}.
	
	\bibitem[{\citenamefont{Dimitriou and McKinley}(2019)}]{DIMITRIOU2019}
	\bibinfo{author}{\bibfnamefont{C.~J.} \bibnamefont{Dimitriou}}
	  \bibnamefont{and} \bibinfo{author}{\bibfnamefont{G.~H.}
		  \bibnamefont{McKinley}}, \bibinfo{journal}{Journal of Non-Newtonian Fluid
		  Mechanics} \textbf{\bibinfo{volume}{265}}, \bibinfo{pages}{116}
	  (\bibinfo{year}{2019}), ISSN \bibinfo{issn}{0377-0257},
	  \urlprefix\url{https://www.sciencedirect.com/science/article/pii/S0377025718301162}.
	
	\bibitem[{\citenamefont{Shu}(2009)}]{shu2009}
	\bibinfo{author}{\bibfnamefont{C.~W.} \bibnamefont{Shu}},
	  \bibinfo{journal}{SIAM Review} \textbf{\bibinfo{volume}{51}},
	  \bibinfo{pages}{82} (\bibinfo{year}{2009}).
	
	\bibitem[{\citenamefont{Kim and Moin}(1985)}]{kim_application_1985}
	\bibinfo{author}{\bibfnamefont{J.}~\bibnamefont{Kim}} \bibnamefont{and}
	  \bibinfo{author}{\bibfnamefont{P.}~\bibnamefont{Moin}},
	  \bibinfo{journal}{Journal of Computational Physics}
	  \textbf{\bibinfo{volume}{59}}, \bibinfo{pages}{308} (\bibinfo{year}{1985}),
	  \urlprefix\url{https://linkinghub.elsevier.com/retrieve/pii/0021999185901482}.
	
	\bibitem[{\citenamefont{Izbassarov et~al.}(2018)\citenamefont{Izbassarov,
			  Rosti, Ardekani, Sarabian, Hormozi, Brandt, and
			  Tammisola}}]{izbassarov_etal_2018a}
	\bibinfo{author}{\bibfnamefont{D.}~\bibnamefont{Izbassarov}},
	  \bibinfo{author}{\bibfnamefont{M.~E.} \bibnamefont{Rosti}},
	  \bibinfo{author}{\bibfnamefont{M.~N.} \bibnamefont{Ardekani}},
	  \bibinfo{author}{\bibfnamefont{M.}~\bibnamefont{Sarabian}},
	  \bibinfo{author}{\bibfnamefont{S.}~\bibnamefont{Hormozi}},
	  \bibinfo{author}{\bibfnamefont{L.}~\bibnamefont{Brandt}}, \bibnamefont{and}
	  \bibinfo{author}{\bibfnamefont{O.}~\bibnamefont{Tammisola}},
	  \bibinfo{journal}{International Journal for Numerical Methods in Fluids}
	  \textbf{\bibinfo{volume}{88}}, \bibinfo{pages}{521} (\bibinfo{year}{2018}).
\end{enumerate}

\pagebreak
\section{Supplementary information}

\setcounter{equation}{0}
\setcounter{figure}{0}
\setcounter{table}{0}
\makeatletter
\renewcommand{\theequation}{S\arabic{equation}}
\renewcommand{\thefigure}{S\arabic{figure}}

\subsection{Scale-by-scale energy balance}
\label{app:spectral-balance}
This section gives a derivation of equation~\cref{eq:energy-transfer} from the main article. Firstly, we perform the Fourier transform of the Navier-Stokes equations to obtain an expression for the turbulent kinetic energy spectrum $\hat{E}(\boldsymbol{\kappa},t) \equiv \frac{1}{2}\rho\langle \boldsymbol{\hat{u}}^{*} \cdot \boldsymbol{\hat{u}}\rangle$, where $(\hat{.})$ denotes the Fourier transform into the spectral space, $\boldsymbol{\kappa}$ denotes the wave vector with a magnitude $\kappa$, and the superscript $*$ denotes the complex conjugate;
\begin{equation}\label{eq:massF}
	\boldsymbol{\kappa} \cdot \boldsymbol{\hat{u}}=0,
\end{equation}
\begin{equation}\label{eq:momF}
	\rho\frac{\mathrm{d} \boldsymbol{\hat{u}}}{\mathrm{d} t} + \boldsymbol{\hat{G}} = -\iota \boldsymbol{\kappa} \hat{p} - \mu_\text{f}{\kappa^{2}}\boldsymbol{\hat{u}} + \boldsymbol{\hat{f}}_\text{inj}+ \boldsymbol{\hat{f}}_\text{evp},
\end{equation}
where $\hat G$ is the Fourier coefficient of the non-linear convective term appearing in~\cref{eq:mom} of the main article, and $\iota$ is the imaginary unit. Similar equations can be obtained for the complex conjugate $\hat {\boldsymbol{u}}^{*}$. When~\cref{eq:momF} is multiplied by $\boldsymbol{\hat{u}}^{*}$, the pressure term $-\iota\boldsymbol{\kappa} \cdot \boldsymbol{\hat{u}}^{*} \hat{p}$ vanishes due to the incompressibility constraint (\cref{eq:massF}), and the viscous term $- \mu_\text{f}{\kappa^{2}}\boldsymbol{\hat{u}}\cdot \boldsymbol{\hat{u}}^{*}$ can be expressed in terms of the kinetic energy; $-2 \mu_\text{f}{\kappa^{2}} \hat{E}$. The same holds when multiplying the momentum equation of $\hat {\boldsymbol{u}}^{*}$ by $\hat {\boldsymbol{u}}$. By summing the two equations for $\hat {\boldsymbol{u}}$ and $\hat {\boldsymbol{u}}^{*}$ and dividing by $2$, we have an expression for the time evolution of turbulent kinetic energy $ \hat{E}(\boldsymbol{\kappa},t)$
\begin{equation}\label{eq:Espec}
	\frac{\mathrm{d} \hat{E}(\boldsymbol{\kappa})}{\mathrm{d} t} =  \hat{T}(\boldsymbol{\kappa}) + \hat{V}(\boldsymbol{\kappa}) + \hat{F}_\text{inj}(\boldsymbol{\kappa})+ \hat{F}_{\text{evp}}(\boldsymbol{\kappa}) ,
\end{equation}
where the terms on the right-hand side represent the following contributions: $\hat{T}= -\frac{1}{2}(\boldsymbol{\hat{G}}\cdot \boldsymbol{\hat{u}}^{*}+\boldsymbol{\hat{G}}^{*}\cdot \boldsymbol{\hat{u}})$ is due to the non-linear convective term, $\hat{V} = - 2 \mu_\text{f}{\kappa^{2}} \hat{E}$ is due to the fluid dissipation term, $\hat{F}_\text{inj}=\frac{1}{2}(\boldsymbol{\hat{f}}_\text{inj}\cdot \boldsymbol{\hat{u}}^{*}+\boldsymbol{\hat{f}}_\text{inj}^{*}\cdot \boldsymbol{\hat{u}})$ is due to the external forcing, and $\hat{F}_{\text{evp}}=\frac{1}{2}(\boldsymbol{\hat{f}_{\text{evp}}}\cdot \boldsymbol{\hat{u}}^{*}+\boldsymbol{\hat{f}_{\text{evp}}^{*}}\cdot \boldsymbol{\hat{u}})$ is due to the  non-Newtonian stress. The one-dimensional energy spectrum $E(\kappa,t)$ can be obtained by isotropically averaging~\cref{eq:Espec} over the sphere of radius $\kappa$ (i.e., $E(\kappa,t)=\iint_{S(\kappa)} \hat{E}(\boldsymbol{\kappa},t) \mathrm{d}S(\kappa)$, where $S(\kappa) $ is the sphere defined by $\boldsymbol\kappa\cdot\boldsymbol\kappa= \kappa^{2}$),
\begin{equation}\label{eq:Espec_avg}
	\frac{\mathrm{d} {E(\kappa)}}{\mathrm{d} t} =  T(\kappa) + V(\kappa) + F_\text{inj}(\kappa)+ F_{\text{evp}}(\kappa).
\end{equation}
where $\frac{\mathrm{d} {E}}{\mathrm{d} t}$ becomes zero for a statistically stationary flow. Integrating \cref{eq:Espec_avg} from $\kappa$ to infinity, we obtain the energy-transfer balance
\begin{equation}
	\label{eq:E_balance}
	0 = \Pi + \mathcal{D}' + \mathcal{F}_\text{inj}+ \mathcal{N}',
\end{equation}
where $\Pi(\kappa) \equiv \int_\kappa^\infty T (\kappa) \, \mathrm{d}\kappa$,\,\, $\mathcal{D}'(\kappa) \equiv \int_\kappa^\infty V(\kappa) \, \mathrm{d}\kappa$,\,\,  $\mathcal{F}_\text{inj}(\kappa) \equiv \int_\kappa^\infty F_\text{inj}(\kappa) \, \mathrm{d}\kappa$, and $\mathcal{N}'(\kappa) \equiv \int_\kappa^\infty F_\mathrm{evp}(\kappa) \, \mathrm{d}\kappa$ represent the contributions to the spectral power balance from the non-linear convective, fluid dissipation, turbulence forcing, and non-Newtonian terms, respectively. The fluid dissipation term can be expressed as $\mathcal{D}(\kappa) = -\int_0^\kappa V(\kappa) \, \mathrm{d}\kappa=\mathcal{D}'(\kappa)+{\langle\epsilon_\text{f}\rangle}$, where ${\langle\epsilon_\text{f}\rangle}=-\int_0^\infty V(\kappa) \, \mathrm{d}\kappa$ is the rate of energy dissipated by the fluid viscosity. Similarly, the non-Newtonian contribution can be written as $\mathcal{N}(\kappa) = -\int_0^\kappa F_{\text{evp}}(\kappa) \, \mathrm{d}\kappa=\mathcal{N}'(\kappa) + {\langle\epsilon_\text{n}\rangle}$, where ${\langle\epsilon_\text{n}\rangle}=-\int_0^\infty F_{\text{evp}}(\kappa)  \, \mathrm{d}\kappa$ is the non-Newtonian dissipation. Substituting these in~\cref{eq:E_balance}, we obtain the energy balance equation (Eq. (1)) used in the main article.

\clearpage
\subsection{Fluid dissipation and non-Newtonian dissipation}
The rate of turbulent kinetic energy dissipated by the fluid viscosity is 
$\epsilon_\text{f}\equiv 2 \mu_\text{f} s_{ij} s_{ij} $, where $i$ and $j$ are indices for the Cartesian components of a tensor, repeated indices are summed over, and 
$s_{ij}\equiv (\partial u_i/\partial x_j + \partial u_j/\partial x_i)/2$ 
is the strain rate.
Analogously, we can define the rate of turbulent kinetic energy dissipated by the non-Newtonian extra stresses $\epsilon_\text{n}\equiv-\boldsymbol{u} \cdot (\nabla \cdot \boldsymbol{\tau}$). When averaged throughout a triperiodic volume in statistically steady state, this can be expressed as $\langle\epsilon_\text{n}\rangle = \left\langle \frac{\mathrm{tr}(\boldsymbol{\tau})}{2\lambda} \max(0,\frac{\tau_{d}-\tau_{y}}{\tau_{d}}) \right\rangle$.

\subsection{Supplementary results}
\begin{figure}[!h]
	\centering
	\includegraphics{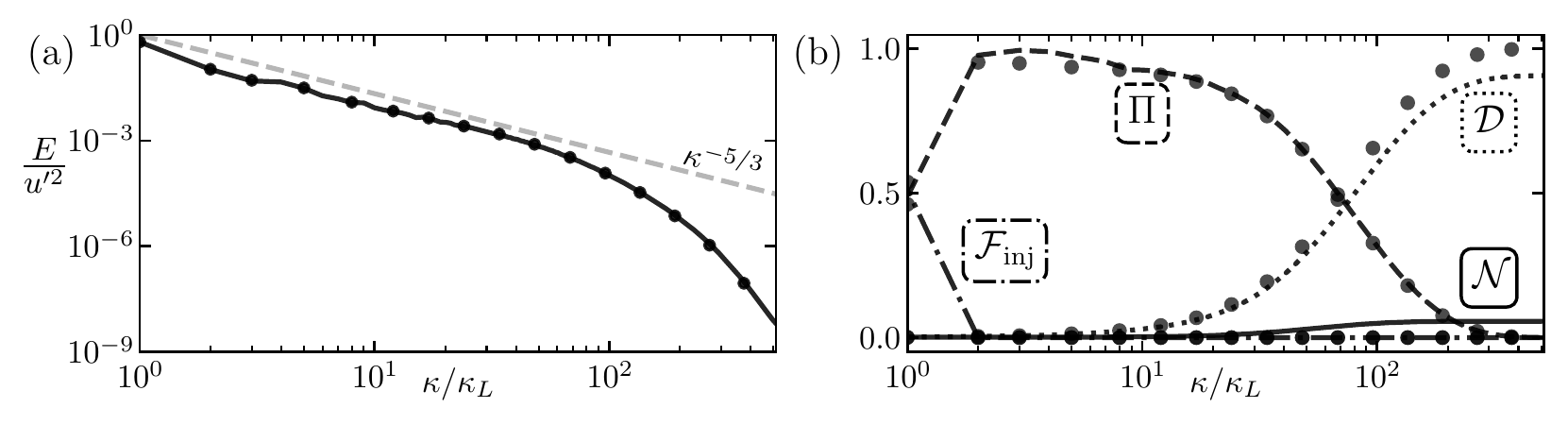}
	\caption {(a) Turbulent kinetic energy spectrum and (b) energy-transfer balance for the Newtonian flow (circles) and $Bi_\Lambda=0$ (lines).}
	\label{fig:Fig1S}
\end{figure}

\begin{figure}[!h]
	\centering
	\includegraphics{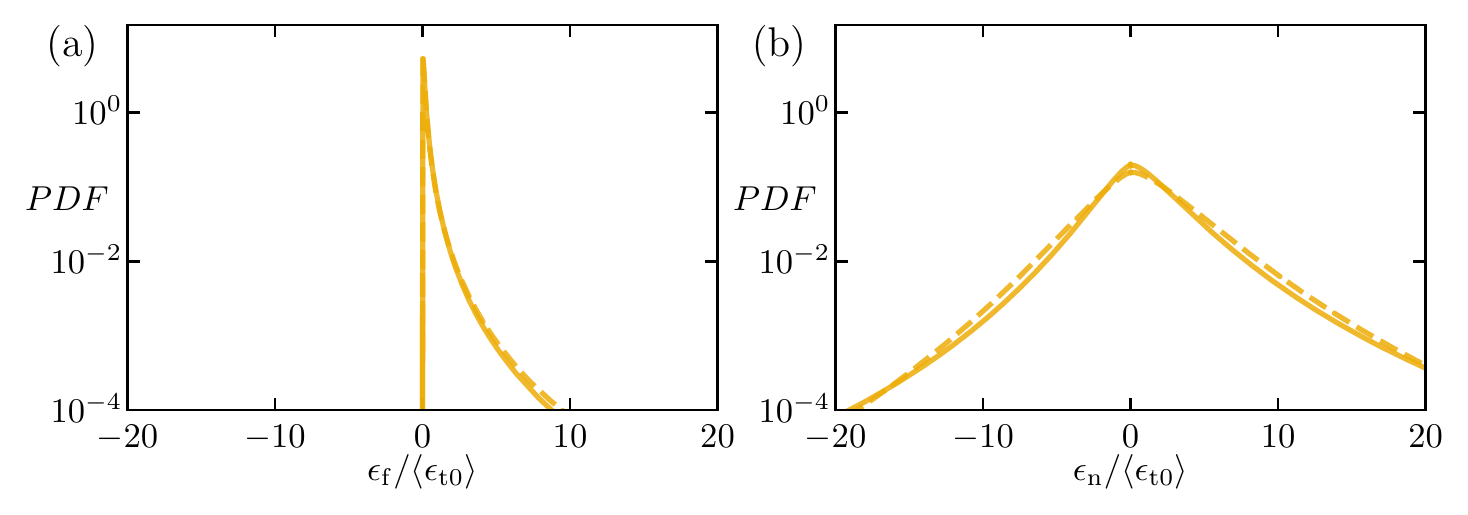}
	\caption{Probability distribution function (PDF) of (a) the fluid dissipation rate $\epsilon _\text{f}$ and of (b) the non-Newtonian dissipation rate $\epsilon _\text{n}$ averaged over time for $Bi_\Lambda =25$ at $Wi_\Lambda = 10^{-3}$ (solid line) and $Wi_\Lambda = 10^{-4}$ (dashed line). Here, the finite value of $\epsilon_\text{n}$ in the limit of $Wi_\Lambda \to 0$ is reminiscent of the dissipative anomaly in Newtonian flows~\cite{S_donzis_scalar_2005}}
	\label{fig:Fig2S}
\end{figure}

In this section, we provide additional results that support our findings explained in the main article.
\cref{fig:Fig1S} shows the good agreement between $Bi_\Lambda=0$ case and the Newtonian case in the energy spectrum (\cref{fig:Fig1S}a) and energy-transfer balance (\cref{fig:Fig1S}b) despite the small contribution of the non-Newtonian stress appearing in the energy balance of the $Bi_\Lambda=0$ case, which is due to the tiny amount of elasticity given to the flow ($Wi_\Lambda= {10^{-3}}$). To further verify that the elastic effect is {negligible} in our {results}, we run simulations at two different Weissenberg numbers  {$Wi_\Lambda = 10^{-3}$ and $Wi_\Lambda = 10^{-4}$} for the highest {Bingham} number considered, i.e., $Bi_\Lambda=25$. \cref{fig:Fig5S} shows the two cases give the same intermittency correction, and \cref{fig:Fig2S} shows the probability distribution functions of the fluid dissipation $\epsilon_\text{f}$ and the non-Newtonian dissipation $\epsilon_\text{n}$ remain unaltered for both cases.

\begin{figure}[!h]
	\centering
	\includegraphics{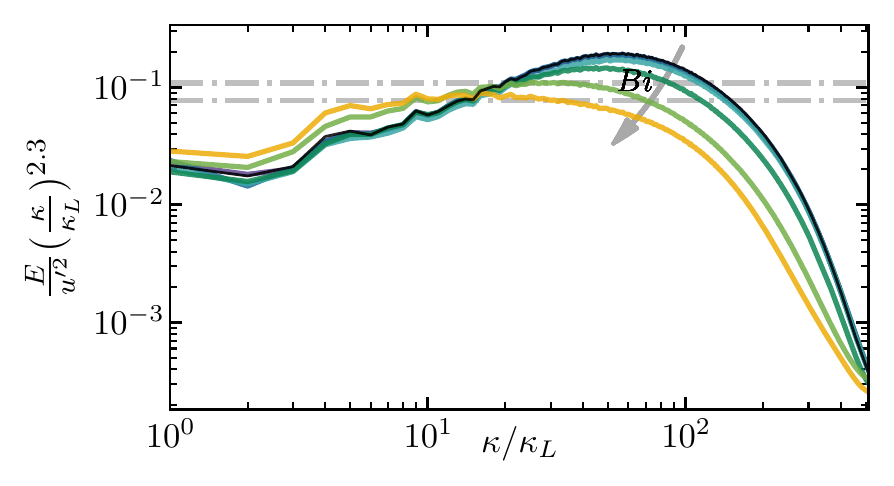}
	\caption{Compensated turbulent kinetic energy spectra ($E \, \kappa^{2.3}$) of EVP flows with various Bingham numbers, plotted using the same representative colours used in Fig. 2 in the main article. The dash-dotted lines represent an apparent $-2.3$ scaling which emerges at high $Bi_\Lambda$.}
	\label{fig:Fig3S}
\end{figure}

The emergence of the new scaling $E\sim\kappa^{-2.3}$ observed in between the small and intermediate scales in the energy cascade for high $Bi_\Lambda$ can be seen more clearly when the spectrum is pre-multiplied by the scaling, as shown in~\cref{fig:Fig3S}. For the two highest Bingham numbers, approximately constant regions emerge where the scaling holds. We can use this new scaling and the fact that the energy spectrum and structure functions form a Fourier transform pair~\cite{S_davidson_turbulence_2015} to predict how the structure functions depend on separation $r$: $S_2\sim r^{1.3},$ $S_3\sim r^{1.95},$ $S_4\sim r^{2.6},$ and $S_6\sim r^{3.9}.$

\begin{figure}[!h]
	\centering
	\includegraphics{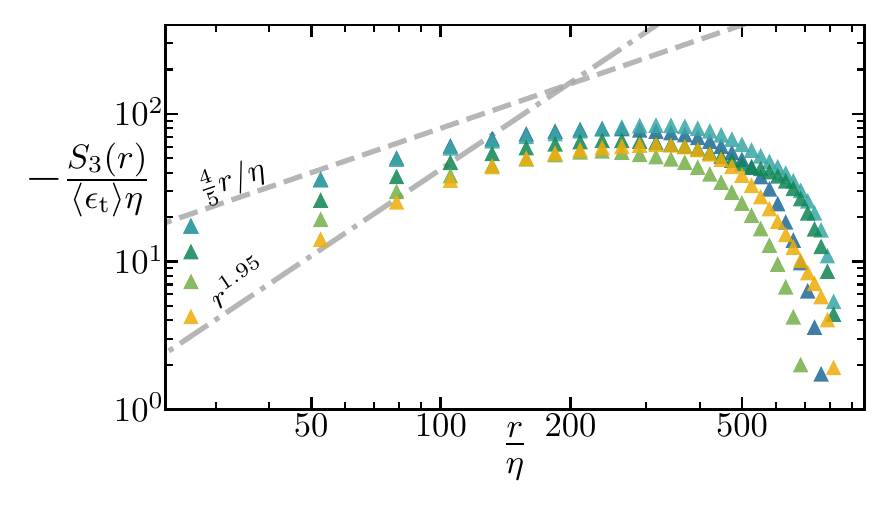}
	\caption{The third order structure functions. The dashed line shows the K41 exact result, while the dash-dotted line represents the expected scaling of $S_3$ using the apparent new non-Newtonian scaling $E\sim \kappa^{-2.3}$.}
	\label{fig:Fig4S_S3ByFluidDissVsR}
\end{figure}

\begin{figure}[!h]
	\centering
	\includegraphics{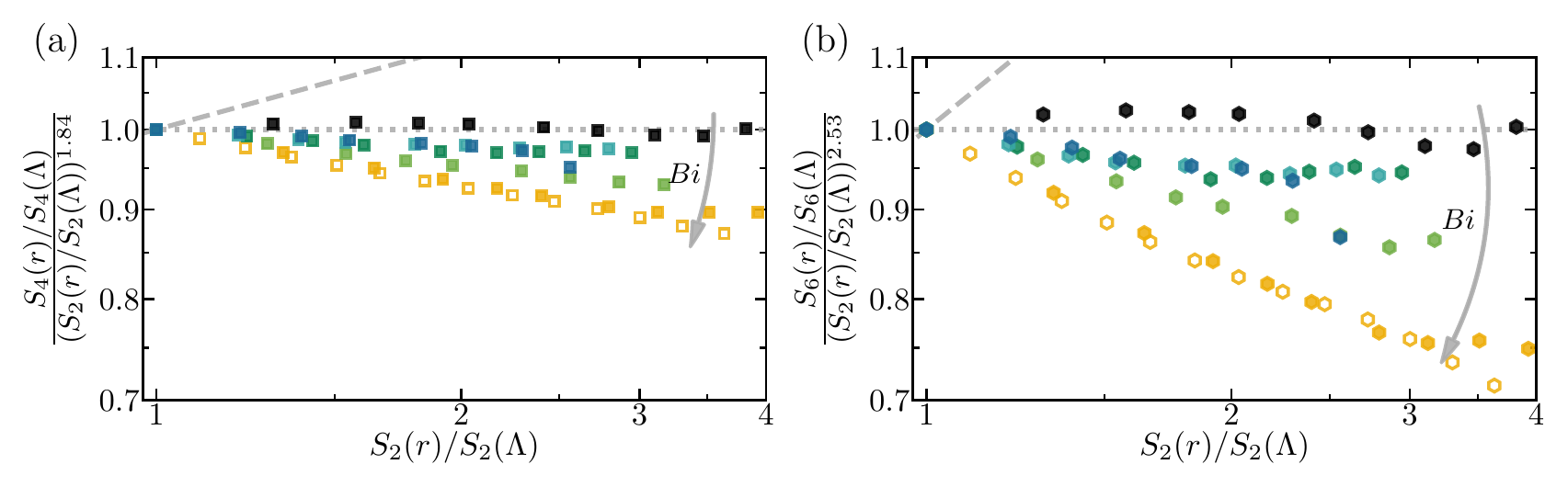}
	\caption{The extended self-similarity form of the structure functions (a) $S_4$ and (b) $S_6$, compensated by the intermittency correction at $Bi_\Lambda =0$.  Filled markers show flows with $Wi_\Lambda = 10^{-3}$ and empty markers show the lower elasticity case $Wi_\Lambda = 10^{-4}$. The dotted line shows the best fit through the $Bi_\Lambda =0$ data, while the dashed line represents the K41 prediction without intermittency.}
	\label{fig:Fig5S}
\end{figure}

The third order structure function is shown in~\cref{fig:Fig4S_S3ByFluidDissVsR}. At low Bingham numbers $S_3$ follows the K41 exact result $S_3=-\frac{4}{5}\langle\epsilon_\text{f}\rangle r$\cite{S_kolmogorov41}. Whereas the $Bi_\Lambda>10$ structure functions support the new scaling $S_3\sim r^{1.95}$ at small scales ($20\eta<r<80\eta$). 
The third order structure function also gives a measure of the direction of turbulent kinetic energy cascade in the flow, $S_3$ negative indicates a direct cascade of energy from large to small scales, whereas $S_3$ positive indicates an inverse cascade~\cite{S_xia_upscale_2011-1,S_cerbus_third-order_2017}.

Finally, in \cref{fig:Fig5S}, we demonstrate further the increased intermittency of the EVP flow due to the fluid plasticity. We do this by showing in the extended self-similarity form the structure functions $S_4$ (\cref{fig:Fig5S}a) and $S_6$ (\cref{fig:Fig5S}b), compensated by the intermittency correction of $Bi_\Lambda= 0$, and plotted against $S_2$. {We can see} clearly how intermittency grows as the fluid becomes more plastic.

\section*{}
	\section{Supplementary references}
	\begin{enumerate}
		\setcounter{enumi}{30} 
		\itemsep0em 
	\expandafter\ifx\csname natexlab\endcsname\relax\def\natexlab#1{#1}\fi
	\expandafter\ifx\csname bibnamefont\endcsname\relax
	\def\bibnamefont#1{#1}\fi
	\expandafter\ifx\csname bibfnamefont\endcsname\relax
	\def\bibfnamefont#1{#1}\fi
	\expandafter\ifx\csname citenamefont\endcsname\relax
	\def\citenamefont#1{#1}\fi
	\expandafter\ifx\csname url\endcsname\relax
	\def\url#1{\texttt{#1}}\fi
	\expandafter\ifx\csname urlprefix\endcsname\relax\def\urlprefix{URL }\fi
	\providecommand{\bibinfo}[2]{#2}
	\providecommand{\eprint}[2][]{\url{#2}}
	
\bibitem[{\citenamefont{Donzis et~al.}(2005)\citenamefont{Donzis, Sreenivasan,
		and Yeung}}]{S_donzis_scalar_2005}
\bibinfo{author}{\bibfnamefont{D.~A.} \bibnamefont{Donzis}},
\bibinfo{author}{\bibfnamefont{K.~R.} \bibnamefont{Sreenivasan}},
\bibnamefont{and} \bibinfo{author}{\bibfnamefont{P.~K.} \bibnamefont{Yeung}},
\bibinfo{journal}{Journal of Fluid Mechanics} \textbf{\bibinfo{volume}{532}},
\bibinfo{pages}{199} (\bibinfo{year}{2005}), ISSN \bibinfo{issn}{0022-1120,
	1469-7645}.

\bibitem[{\citenamefont{Davidson}(2015)}]{S_davidson_turbulence_2015}
\bibinfo{author}{\bibfnamefont{P.}~\bibnamefont{Davidson}},
\emph{\bibinfo{title}{Turbulence: {{An Introduction}} for {{Scientists}} and
			{{Engineers}}}} (\bibinfo{publisher}{{Oxford University Press}},
\bibinfo{year}{2015}), ISBN \bibinfo{isbn}{978-0-19-872259-5}.
	
\bibitem[{\citenamefont{{Kolmogorov}}(1941)}]{S_kolmogorov41}
\bibinfo{author}{\bibfnamefont{A.}~\bibnamefont{{Kolmogorov}}},
\bibinfo{journal}{Akademiia Nauk SSSR Doklady} \textbf{\bibinfo{volume}{30}},
\bibinfo{pages}{301} (\bibinfo{year}{1941}).

\bibitem[{\citenamefont{Xia et~al.}(2011)\citenamefont{Xia, Byrne, Falkovich,
		and Shats}}]{S_xia_upscale_2011-1}
\bibinfo{author}{\bibfnamefont{H.}~\bibnamefont{Xia}},
\bibinfo{author}{\bibfnamefont{D.}~\bibnamefont{Byrne}},
\bibinfo{author}{\bibfnamefont{G.}~\bibnamefont{Falkovich}},
\bibnamefont{and} \bibinfo{author}{\bibfnamefont{M.}~\bibnamefont{Shats}},
\bibinfo{journal}{Nature Physics} \textbf{\bibinfo{volume}{7}},
\bibinfo{pages}{321} (\bibinfo{year}{2011}), ISSN \bibinfo{issn}{1745-2473,
	1745-2481}.

\bibitem[{\citenamefont{Cerbus and
		Chakraborty}(2017)}]{S_cerbus_third-order_2017}
\bibinfo{author}{\bibfnamefont{R.~T.} \bibnamefont{Cerbus}} \bibnamefont{and}
\bibinfo{author}{\bibfnamefont{P.}~\bibnamefont{Chakraborty}},
\bibinfo{journal}{Physics of Fluids} \textbf{\bibinfo{volume}{29}},
\bibinfo{pages}{111110} (\bibinfo{year}{2017}), ISSN
\bibinfo{issn}{1070-6631}.

\end{enumerate}
\end{document}